\documentclass[aps,superscriptaddress]{revtex4}
\usepackage{epsfig}
\usepackage{amsfonts,amssymb,amsmath}

\newcommand{\smallfrac}[2][1]{\mbox{$\textstyle \frac{#1}{#2}$}}
\newcommand{\Tr}{\mathrm{Tr}}

\renewcommand{\epsilon}{\varepsilon}

\begin{document}

\title{Local Realism of Macroscopic Correlations}
\date{\today}
\author{R. \surname{Ramanathan}}
\affiliation{Centre for Quantum Technologies, National University of Singapore, 3 Science Drive 2, 117543 Singapore, Singapore} 
\author{T. \surname{Paterek}}
\affiliation{Centre for Quantum Technologies, National University of Singapore, 3 Science Drive 2, 117543 Singapore, Singapore}
\author{A. \surname{Kay}}
\affiliation{Centre for Quantum Technologies, National University of Singapore, 3 Science Drive 2, 117543 Singapore, Singapore}
\affiliation{Keble College, Parks Road, Oxford, OX1 3PG, United Kingdom}
\author{P. \surname{Kurzy\'nski}}
\affiliation{Centre for Quantum Technologies, National University of Singapore, 3 Science Drive 2, 117543 Singapore, Singapore}
\affiliation{Faculty of Physics, Adam Mickiewicz University, Umultowska 85, 61-614 Pozna\'{n}, Poland} 
\author{D. \surname{Kaszlikowski}}
\email{phykd@nus.edu.sg}
\affiliation{Centre for Quantum Technologies, National University of Singapore, 3 Science Drive 2, 117543 Singapore, Singapore}
\affiliation{Department of Physics, National University of Singapore, 2 Science Drive 3, 117542 Singapore, Singapore}
\begin{abstract}
We show that for macroscopic measurements which cannot reveal full information about microscopic states of the system,
the monogamy of Bell inequality violations present in quantum mechanics 
implies that practically all correlations between macroscopic measurements can be described by local realistic models. 
Our results hold for sharp measurement and arbitrary closed quantum systems.
\end{abstract}
\maketitle

\section{Introduction} 

The notion of local realism posits that measurable properties of physical systems exist before measurements are performed, and that relativistic causality holds. This point of view was brought to the attention of the physics community in the famous paper by Einstein, Podolsky and Rosen in 1935 \cite{epr}, where it was argued that quantum mechanics is an incomplete theory in need of further refinement to bring it in line with local realism. In 1964, John Bell used astonishingly simple reasoning in the form of an algebraic inequality (Bell inequality) to demonstrate that local realism is in contradistinction with the predictions of quantum theory \cite{bell64}. His findings have been confirmed in numerous experiments in which various loopholes, which potentially still allow a local realistic description of the measured data, were closed individually \cite{loopholes1,loopholes2,loopholes3,loopholes4,loopholes5,loopholes6}. Although there is still no conclusive experiment closing all the loopholes at the same time, most scientists think that on the microscopic scale the world is {\it not} local realistic.

The macroscopic world we experience, to the contrary, is described by classical physics; a local realistic theory. One of the most fundamental questions one can ask is how a local realistic macroscopic world emerges from the microscopic scale, on which level it cannot be described by local realism. A number of resolutions to this question have been suggested. The more radical ones, the so-called collapse models \cite{collapse1,collapse2,collapse3,collapse4,collapse5,collapse6}, predict that quantum mechanics will fail for sufficiently complex systems. Another approach is to look for classicality as a limit of quantum phenomena. The decoherence programme derives the lack of superposition of the pointer state of the measuring apparatus from an inevitable interaction between the quantum system and its environment, see for instance Ref.\ \cite{decoh1,decoh2,decoh3}. A conceptually different approach focuses on the limits of observability of quantum effects in macroscopic objects \cite{coarse-graining1,coarse-graining2,coarse-graining3,coarse-graining4}.

The steady progress in experimental techniques allows one to perform measurements that were considered infeasible decades ago. Experiments have reached a level of sophistication where several spins can be manipulated coherently for sufficiently long times to perform small quantum computations \cite{nmr}. In spite of this tremendous progress, one still faces a formidable challenge to manipulate systems consisting of a macroscopic number, perhaps of the order of $10^{23}$, of particles. Although one cannot exclude such a possibility in the future, at the present moment it is simply an experimental impossibility. 

The purpose of this paper is to consider the nature of the correlations that we can reasonably measure on these macroscopic systems with existing experimental capabilities, showing that if the number of measured particles is large enough, a local realistic description emerges, regardless of the quantum state of the system. The intuition behind this result is that macroscopic measurements do not reveal the properties of individual particles, and quantum correlations are monogamous \cite{mono1,mono2,mono3,mono4,mono5,mono6,mono7} while the classical correlations are not.  In fact, we provide answers to two subtly different questions, necessitating two different approaches. These two questions are, in effect ``Why does nature appear classical in the macroscopic limit?" and ``Why does quantum mechanics appear classical in the macroscopic limit?". The distinction arises because, to date, while we have overwhelming confirmation of quantum mechanics for the sets of observables that we can access in the lab, there are certainly correlation functions of macroscopically large systems, such as the systems we will examine in this paper, that have never been tested for conformance to quantum mechanics; it is possible that nature functions quantum mechanically in experimentally accessible observables, but behaves differently on these scales. Hence, we make a distinction, although we will see that it does not significantly affect the conclusions.

\section{Macroscopic measurements and LHV Models}

In order to precisely define our concept of macroscopic measurements, consider systems of many qubits. This is often a good approximation to systems composed of magnetic materials and metals \cite{ashcroft}. We are interested in experimentally feasible measurements performed on macroscopic regions of the system whose results can be known with arbitrary precision. The simplest measurement of this kind is magnetization along some direction, which is the average projection of all spins on the given direction. The outcome of this measurement does not reveal information about the spin projections of individual particles; there are many configurations of individual spin projections that give the same magnetization. The situation is therefore analogous to statistical mechanics, where one macrostate is realized by the averaging over an enormous number of microstates. 

Magnetization observables are described by one-body operators that can be written as $\sum_{k}\vec{n}\cdot\vec{\sigma}_k$, where $\vec{\sigma}_k=(\sigma_x,\sigma_y,\sigma_z)$ are the standard Pauli operators acting on the $k$th particle and ${\vec n}$ is a 3 component vector of unit length. One could also consider $M$-body observables that read $\sum_{\kappa}O_{\kappa},$ where $\kappa$ contains all different subsets of $M$ particles and $O_{\kappa}$ is an arbitrary $M$-qubit Hermitian operator. These are increasingly hard to implement experimentally with increasing $M$ (or to extract from the measurement results of single-body operators, as would be the case with the variance, which is a two-body operator).
For this reason, we focus on magnetization measurements as the most feasible scenario
and later we extend our considerations to the case of $M$-body measurements to show that 
they do not change the central thesis of this paper, up to some high $M$ threshold.

\begin{figure}
\begin{center}
\includegraphics[scale=0.4]{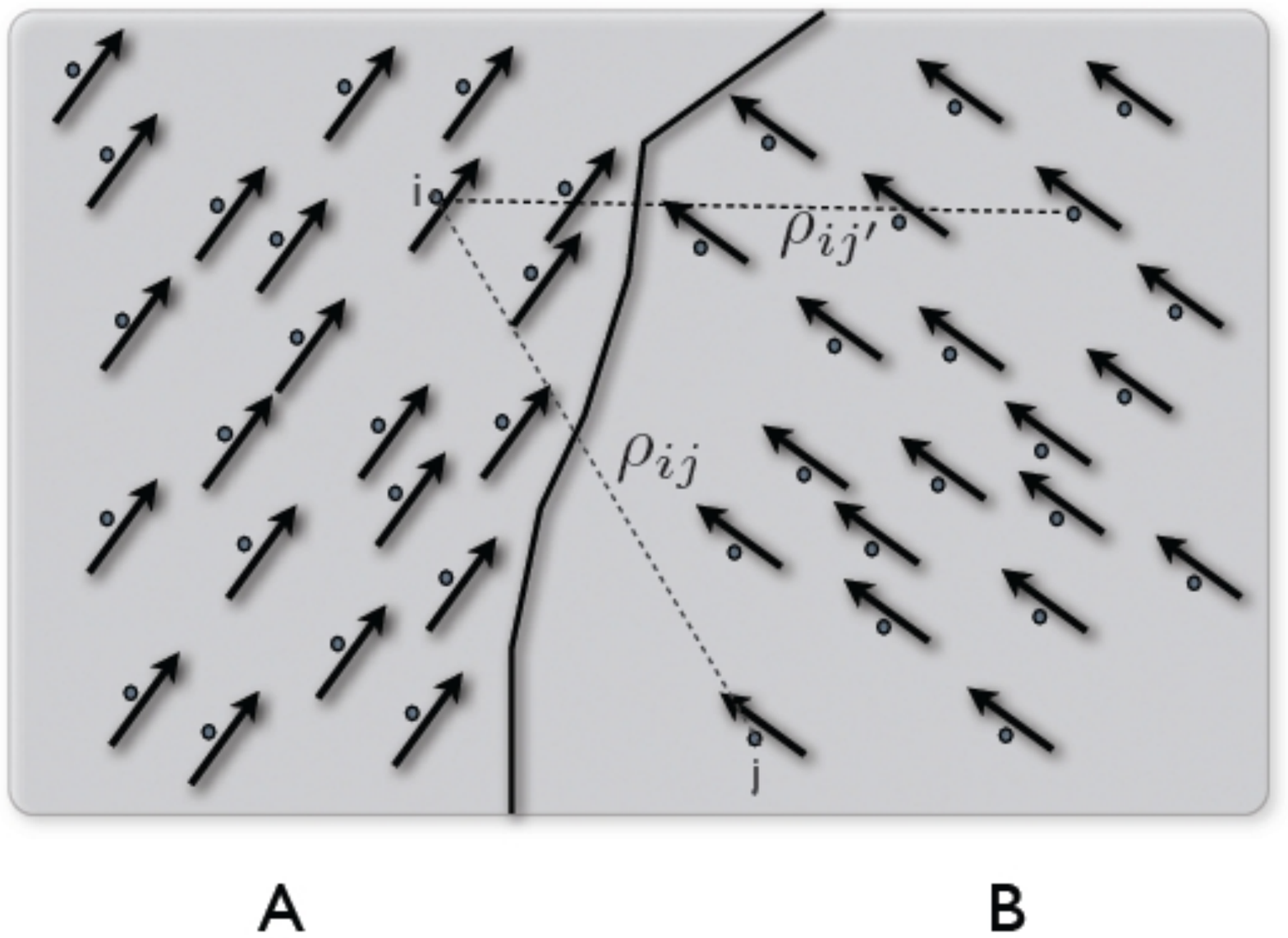}
\end{center}
\vspace{-6cm}
\caption{Measurements of local magnetizations on a macroscopic object. 
The arrows represent directions of magnetization measurements in macroscopic regions $A$ and $B$.
Each microscopic qubit in a given region is measured along the same direction. The correlations between magnetization measurements are determined by the effective two-qubit state $\rho_{eff}^{AB}$ being the uniform average over all reduced density matrices $\rho_{ij}$. The monogamous nature of quantum correlations limits the strength of correlations contained in the states $\rho_{ij}$ and $\rho_{ij'}$ because they share the common qubit $i$.}
 \label{fig1}
\end{figure}

We investigate a lattice of macroscopically many qubits, $N\approx 10^{23}$, prepared in some state $\rho$,
and will prove the existence of a local hidden variable model for the correlations between magnetization measurements on macroscopic regions of these qubits.
As an illustration, consider dividing the lattice into two disjoint regions $A$ and $B$, as depicted in Fig.\ \ref{fig1}, containing $N_A,N_B$ qubits respectively, where $N_A,N_B$ are of the order of $N$. In each of the regions, we perform a measurement of local magnetizations $\mathcal{M}_{\vec a}$ ($\mathcal{M}_{\vec b}$) along some directions $\vec{a}$ ($\vec{b}$):
\begin{equation}
\mathcal{M}_{\vec{a}} \equiv \sum_{i \in A} \vec{a} \cdot \vec{\sigma}_{i} \quad \textrm{ and } \quad \mathcal{M}_{\vec{b}} \equiv \sum_{j \in B} \vec{b} \cdot \vec{\sigma}_{j}.
\end{equation}
Quantum correlations between the magnetization measurements in the state $\rho$,
$\mathbb{E}_{\vec a \vec b} = \langle \mathcal{M}_{\vec{a}} \otimes \mathcal{M}_{\vec{b}} \rangle_{\rho}$,
 are given by
the sum of microscopic correlations between all pairs of qubits from different regions:
\begin{eqnarray}
\mathbb{E}_{\vec a \vec b} &=& \sum_{i\in A} \sum_{j\in B}  \Tr\left((\vec{a} \cdot \vec{\sigma}_i \otimes \vec{b} \cdot \vec{\sigma}_j )\rho\right).
\end{eqnarray}
Since the very same measurements are performed on all the microscopic pairs,
the macroscopic magnetization correlations are effectively described by the averaged state of two qubits:
\begin{equation}
\mathbb{E}_{\vec a \vec b} = N_A N_B \Tr \left((\vec{a} \cdot \vec{\sigma} \otimes \vec{b} \cdot\vec{\sigma}) \rho_{eff}^{AB}\right),
\label{trick}
\end{equation}
where the effective two-qubit state is described by the positive semi-definite operator
\begin{equation}
\rho_{eff}^{AB}=\frac{1}{N_AN_B} \sum_{i \in A} \sum_{j \in B}\rho_{ij},
\end{equation}
and $\rho_{ij}$ is the reduced density matrix for $i$th qubit at $A$ and $j$th qubit at $B$. These different expectation values can then be combined together using coefficients $\alpha({\vec a},{\vec b})$ for the $S_A,S_B$ different measurement settings on Alice's and Bob's partitions respectively, to give what we refer to as a macroscopic Bell parameter:
$$
\langle{\mathcal B}\rangle=\sum_{{\vec a},{\vec b}}\alpha({\vec a},{\vec b})\mathbb{E}_{\vec a \vec b}.
$$

A set of correlations $\mathcal{E}_{\vec a \vec b}$ admits a local hidden variable model (LHV) if there are parameters $\lambda$, distributed with probability density $\mu(\lambda)$, and local response functions $J_A(\vec{a},\lambda)$ and $J_B(\vec{b},\lambda)$ such that 
\begin{eqnarray}
\mathcal{E}_{\vec a \vec b} &=& \int\ d\lambda \mu(\lambda)J_A(\vec{a},\lambda)J_B(\vec{b},\lambda).
\label{lhv}
\end{eqnarray}   

Applied to our scenario a set of quantum correlations $\mathbb{E}_{\vec a \vec b}$ admits an LHV model
as soon as we can construct such a model for correlations obtained from the effective state $\rho_{eff}^{AB}$.
Therefore, $\rho_{eff}^{AB}$ will always be the focus of our study.

Note that whatever results we succeed in deriving regarding a set of states $\rho$ which do not violate some class of macroscopic Bell inequalities, will apply equally to the class of states $(\prod_{i=1}^{N_A}U_i^A\prod_{j=1}^{N_B}U_j^B)\rho(\prod_{i=1}^{N_A}U_i^A\prod_{j=1}^{N_B}U_j^B)^\dagger$ for measurement settings $U_i^A(\vec{a}\cdot{\vec \sigma}){U_i^A}^\dagger$ and $U_i^B(\vec{b}\cdot{\vec \sigma}){U_i^B}^\dagger$ under the same conditions. For instance, if we prove that no state $\rho$ violates a set of Bell inequalities, this instantly generalizes to Bell inequalities which allow for some variation of magnetic fields over the sample.

\section{Explicit models and quantum complementarity} \label{sec:sing}

In this section, we will show that $\rho_{eff}^{AB}$ admits LHV description for two magnetization measurements performed on up to $\log_2 N$ macroscopic regions.
Our proof will proceed by utilizing the quantum character of the magnetization measurements, but makes no assumption on the behavior of correlations which do not enter magnetization outcomes (which may have never been proven to behave quantumly).

A set of four correlations measured on a two-qubit state with one of two local observables admits an explicit LHV model of Ref. \cite{zukowski}
if elements of the correlation tensor $T_{kl} = \Tr(\sigma_k \otimes \sigma_l \rho)$
satisfy \cite{zukowski}:
\begin{eqnarray}
\mathcal{L} \equiv \sum_{k,l=x,y}T_{kl}^2\leq 1,
\label{ZUKBRUK}
\end{eqnarray}
where orthogonal local directions $x$ and $y$ are defined to be along sum and difference of the two local setting vectors.
Note that this condition does not require orthogonal measurement settings in a Bell experiment.

We apply this condition to the effective state $\rho_{eff}^{AB}$.
The elements of its correlation tensor read
\begin{equation}
T_{kl} = \frac{1}{N_A N_B} \sum_{i \in A} \sum_{j \in B} T_{kl}^{(ij)},
\end{equation}
where $T_{kl}^{(ij)}$ is the component of the correlation tensor for particles $i$ and $j$ in the regions $A$ and $B$ respectively.
Substituting into (\ref{ZUKBRUK}) gives
\begin{equation}
\mathcal{L} = \frac{1}{N_A^2 N_B^2} \sum_{i,i' \in A} \sum_{j,j' \in B} \vec T^{(ij)} \cdot \vec T^{(i'j')},
\end{equation}
with $\vec T^{(ij)} = (T_{xx}^{(ij)},T_{xy}^{(ij)},T_{yx}^{(ij)},T_{yy}^{(ij)})$.
In the next step, we write $\mathcal{L}$ as a combination of vectors $\vec P^{(ij)} = (T_{xx}^{(ij)},T_{xy}^{(ij)},T_{yx}^{(i(j+1))},T_{yy}^{(i(j+1))})$
and $\vec Q^{(ij)} = (T_{xx}^{(i(j+1))},T_{xy}^{(i(j+1))},T_{yx}^{(ij)},T_{yy}^{(ij)})$,
the components of which are expectation values of mutually anti-commuting operators:
\begin{equation}
\mathcal{L} = \frac{1}{2 N_A^2 N_B^2} \sum_{i,i' \in A} \sum_{j,j' \in B} \Big( \vec P^{(ij)} \cdot \vec P^{(i'j')} + \vec Q^{(ij)} \cdot \vec Q^{(i'j')} \Big).
\end{equation}
The components of vectors $\vec P^{(ij)}$ and $\vec Q^{(ij)}$ involve correlations between two pairs of micro-systems,
pair $(ij)$ and $(i(j+1))$, i.e. the sum is modulo $N_B$.
The monogamous nature of correlations between these pairs, which stem from quantum complementarity, limits the lengths of
$\vec P^{(ij)}$ and $\vec Q^{(ij)}$ below one \cite{mono7}, and consequently gives $\mathcal{L} \le 1$.
Thus we have shown that the correlations between local magnetizations in the system of $N$ qubits are of classical nature as long as $N_A$ or $N_B$ is greater than one. In effect, the quantum correlations get diluted in the effective state $\rho_{eff}^{AB}$ due to monogamy between the different pairwise terms, which themselves arise because the observables see the whole quantum state $\rho$ as an equal average over all possible pairs of qubits between regions $A$ and $B$.

We generalize this method to the scenario where the system of $N$ qubits is partitioned into $K$ regions such that there are $N_k$, of the order of $\frac{N}{K}$, particles in each region, with $k=1,\dots K$. 
We prove in Appendix A that when $K \leq \log_2(N)$ there is always an LHV description for all quantum states $\rho$. We stress that the bound on $K$ may not be tight and even more macroscopic observers may still not be able to violate a Bell inequality.

The method can also be extended to the scenario where one measures $M$-body observables (for example, magnetization is a $1$-body observable and magnetic susceptibility is a $2$-body observable), and consequently considers Bell inequalities of $2M$-qubit correlation functions. It can be shown using the above methods that in particular CHSH-like inequalities are not violated by macroscopic systems up to some high threshold $M$. 

\section{Bell monogamy and local realistic macroscopic correlations} \label{sec:AKversion}

Now we prove more general results using the stronger assumption that
quantum predictions are valid even for experiments that cannot be performed in practice.
In practice, each microscopic constituent of a macroscopic system cannot be addressed,
but we assume that the predictions of quantum mechanics hold true even if they could be addressed.
Our approach closely follows the proof technique in \cite{mono5} which proved the monogamy of Bell inequalities. Given the similar nature of proofs in \cite{mono6}, one expects that the results can be extended to discuss why general no-signalling theories would also appear classical, not merely limited to quantum mechanics. However, we have not formalised this expectation.

Our additional assumption about the applicability of quantum mechanics permits us to consider a much more general scenario than the previous section. Let us take a sample of $N$ spins of arbitrary local Hilbert space dimension. The vector of matrices ${\vec \sigma}$ provides a Hermitian basis for the operators in the local Hilbert space. Partition these spins into $k$ partitions of $N_A$, $N_B\ldots N_K$ particles respectively. On each of these partitions $X$, we will be able to choose from $S_X$ measurement settings. A given measurement setting will be denoted by $i_X$, and the corresponding measurement outcome by $j_X$. We can denote the different measurement operators by $E^X_{i_X,j_X}$, which are POVM elements for a given measurement setting and outcome. They satisfy a completeness relation $\sum_j E^X_{i,j}= \openone $. So, we can write a rather generic Bell inequality (which includes those previously defined as a subset) in terms of
$$
\langle{\mathcal B}\rangle=\sum_{{\vec i},{\vec j}}\alpha({\vec i},{\vec j})\Tr\left(\rho_{eff}^{AB\ldots K}(E^{A}_{i_A,j_A}\otimes E^{B}_{i_B,j_B}\otimes\ldots E^{K}_{i_K,j_K})\right)
$$
where ${\vec i}$ is a vector of the measurement settings $i_A\ldots i_K$ and $\rho_{eff}^{AB\ldots K}$ is similarly defined to $\rho_{eff}^{AB}$, i.e.,
$$
\rho_{eff}^{AB\ldots K}=\frac{1}{N_AN_B\ldots N_K}\sum_{a\in A\ldots k\in K}\rho_{a,b\ldots k}.
$$
Now we will show that the following quantum probability distribution 
$$
p({\vec i},{\vec j})=\Tr\left(\rho_{eff}^{AB\ldots K}(E^{A}_{i_A,j_A}\otimes E^{B}_{i_B,j_B}\otimes\ldots E^{K}_{i_K,j_K})\right),
$$
admits a LHV model. This can be done provided the number of measurement settings, $S_X$, is equal to the number of spins in the partition, $N_X$ for all $X\in\{A, B\ldots K\}$. To start, we define vectors ${\vec m_X}$ of $S_X$ elements, which read like a script for a deterministic protocol of what measurement results to give provided with a measurement setting: if the measurement setting is $i_X$, element $m_X^{i_X}$ is what should be given as outcome $j_X$. With this in place, we are in a position of give the LHV strategy -- a source of shared randomness between all the parties selects a set of vectors ${\vec m_A}, {\vec m_B}\ldots{\vec m_K}$ with probability
\begin{equation}
\Tr\left(\rho'(E^A_{\vec m_A}\otimes E^B_{\vec m_B}\otimes\ldots E^K_{\vec m_K})\right)	\label{eqn:physicalassumption}
\end{equation}
where $\rho'$ is any quantum state that has every $k$-qubit reduced density matrix drawn from one Alice, one Bob etc.\ is equal to $\rho_{eff}^{AB\ldots K}$ and where
$$
E^A_{\vec m_A}=E^A_{1,m_A^1}\otimes E^A_{2,m_A^2}\otimes \ldots E^A_{S_A,m_A^{S_A}}
$$
(This is well defined if $S_A=N_A$). Having jointly selected these vectors, then the parties wait until they're told what their measurement setting $i_X$ is, at which point they give the outcome $m_X^{i_X}$. If we use this strategy, the resultant probability distribution is
$$
p({\vec i},{\vec j})=\sum_{{\vec m_A}\ldots{\vec m_K}}\Tr(\rho'E^A_{\vec m_A}\otimes E^B_{\vec m_B}\otimes\ldots E^K_{\vec m_K})\delta_{m^{i_A}_A,j_A}\ldots\delta_{m^{i_K}_K,j_K},
$$
which you will readily see is equal to the desired distribution by using the completeness relations of the POVM operators. So, this will lead us to conclude that if at least one example of a state $\rho'$ exists, for a given $\rho_{eff}^{AB\ldots K}$, then the original state $\rho$ cannot violate a macroscopic Bell inequality of $S_A=N_A$, $S_B=N_B\ldots$ settings. However, we can always construct $\rho'$ from $\rho$. Let $\Pi_X$ be a permutation over all spins of a given partition $X$. Thus,
$$
\rho'=\frac{1}{|\Pi_A|\ldots |\Pi_K|}\sum_{\Pi_A\ldots \Pi_K}(\Pi_A\otimes\Pi_B\ldots\otimes \Pi_K)\rho(\Pi_A\otimes\Pi_B\ldots\otimes \Pi_K)^\dagger.
$$

The result readily extends in two ways. Firstly, observe that in the $N_A$ measurement settings (for instance), any two can be set equal to each other, and the result still holds. Thus, in fact, the result holds provided all $S_X\leq N_X$. Secondly, we can examine many-body observables. For $M$-body observables, we can redefine the effective Bell inequality of $\rho_{eff}$ to be over $M$ physical spins in each partition (although this requires that those $M$-body observables can be applied to all possible subsets of $M$ spins, whereas one might prefer to impose a locality constraint). This has the knock-on effect of simply rescaling the limiting number of Bell measurements to $N_X/M$, assuming this is an integer. So, a system of say $10^{23}$ particles divided into $10^7$ partitions, and involving $10^7$-body observables would still require at least $10^9$ measurement settings to possibly measure some violation of a Bell inequality, which we consider infeasible.

In order to reach this result, we were required to calculate the probabilities in Eqn.\ (\ref{eqn:physicalassumption}), which are probabilities defined beyond the limit up to which quantum mechanics has been tested. This is why we have made the distinction between the two derivations; that of the previous section did not require this assumption. However, in some sense, this is not required here either. The probabilities of Eqn.\ (\ref{eqn:physicalassumption}) can be just that, probabilities devoid of further physical interpretation. Thus, even though the physical specimen that we are measuring may not be in the quantum state $\rho$ and these expectation values do not actually exist, all we need to know is that the set of measurements that we can perform appear quantum mechanical. If they indeed appear so, they must appear as if they were originating from some quantum state $\rho$ (if there were no quantum state $\rho$ compatible with all the measurement results, we would conclude that the system is not behaving quantumly). Thus, we can use the mathematical formalism of quantum theory to manipulate this hypothetical $\rho$ and give us the LHV.

This no-go theorem gives a very strong bound on the degree of control we would need over large systems for there to possibly be a violation of a Bell inequality. Indeed, it is quite tight since it says that for two parties with $N_A=1$ and $N_B=1,2$ with $S_A=S_B=1,2$ there cannot be a Bell violation, whereas one can show that there is a violation for $N_A=1$ and $N_B=1,2$ with $S_A=S_B=2,3$ (the $N_B=1$ case is just CHSH. The $N_B=2$ case uses a 3-setting Bell inequality found in \cite{gisin}). Another interesting feature, however, is that there are some classes of states which we can show will never violate these macroscopic Bell inequalities, no matter how many measurement settings are allowed in the Bell inequality, as we will see in the following section.

\section{Rotationally invariant systems}

Stronger results can be proved for restricted classes of $N$-qubit states $\rho$, such as those which are rotationally invariant, i.e., 
\begin{eqnarray}
&&\rho = U^{\otimes N}\rho(U^{\otimes N})^{\dagger},
\label{rot}
\end{eqnarray}
for all single qubit unitaries $U$. This is a wide class of physically important states such as thermal states of the Heisenberg model.

First of all, we notice that any reduced density matrix $\rho_{ij}$ obtained from the density matrix $\rho$ satisfying the relation (\ref{rot}) is rotationally invariant, i.e., $\rho_{ij}=U\otimes U\rho_{ij}U^{\dagger}\otimes U^{\dagger} = V_{ij} |\psi_-\rangle\langle\psi_-|_{ij}+(1-V)\frac{\openone_i \otimes \openone_j}{4}$ \cite{werner}. Thus, the effective state $\rho_{eff}^{AB}$ inherits the same property:
\begin{eqnarray}
&&\rho_{eff}^{AB}  = V|\psi_-\rangle\langle\psi_-|_{AB}+(1-V)\frac{\openone_A\otimes \openone_B}{4},
\end{eqnarray} 
where $-\frac{1}{3}\leq V\leq 1$. It was proven in Ref. \cite{acin} that for $-\frac{1}{3}\leq V\leq 0.66$ this state admits a LHV description for all sets of projective quantum measurements. The upper bound on this range can be extended to $2/3$ by invoking the results of \cite{tonge} within the formalism presented in \cite{acin}. It is also known \cite{barrett} that if $p\leq\smallfrac[5]{12}$, there is no Bell inequality violation at all, even allowing for POVMs. From our prior description of $\rho_{eff}^{AB}$, we can say that
$$
V=\frac{1}{N_AN_B}\sum_{ij}V_{ij}
$$
and, from singlet monogamy \cite{rvb}, one can prove that
$$
V\leq\frac{R_{ab}+2}{3R_{ab}}
$$
where $R_{ab}=\max(N_A,N_B)$. Thus, provided our sample contains more than two qubits, we can never violate a macroscopic Bell inequality (of any number of settings) composed of projective measurements. If $N_A$ or $N_B\geq 8$, there are no Bell inequalities whatsoever that can be violated. 

\section{Conclusions}  We have studied the conditions under which one can sustain a local realistic description of correlations between macroscopic measurements. We focused on a large system of spins ($N\approx 10^{23}$) in an arbitrary quantum state. The system was partitioned into $k\geq 2$ regions, each containing a number of qubits of the order of $\frac{N}{k}$. In each region, a measurement of magnetizations in several randomly chosen directions was considered. 

We concluded from Sec.\ \ref{sec:sing} that for two-setting Bell inequalities on a total of $N$ qubits divided into two partitions, where each setting is just a local magnetic field direction across all spins of a partition, nature (which could possibly contain some post-quantum correlations which are hidden from the measurements that we can directly make) admits a LHV description. 
In the appendices, we will justify that these results continue to hold when we further divide the partitioning such that there are up to $\log_2 (N)$ parties, and for many-body observables, although the exact threshold for the extent of these many-body operators will be presented in a subsequent publication.  

In contrast, in Sec.\ \ref{sec:AKversion}, we saw directly the possible trade-off between number of parties, extent of the many-body interactions and number of measurement settings traded off, at the expense of having to assume that quantum mechanics is valid beyond where it has been experimentally tested. This trade-off is that if any party can utilise a number of settings which is greater than the ratio of the number of particles in the partition to the extent of the many-body correlations measured, a Bell inequality can potentially be violated. However, given the huge number of particles involved in real systems, implementing this requires a thoroughly absurd experiment. Thus, quantum mechanics appears to produce classical correlations in the macroscopic limit as a result of our experimental limitations.

When viewed as a no-go theorem for the visibility of quantum correlations over and above classical correlations, the interesting direction for future study is to consider sets of measurements which could be implemented but are not covered by the proofs here. It would also be interesting to see if we can construct results which involve many-body operators which are necessarily local on some underlying lattice.

\section{Acknowledgements}
This research is supported by the National Research Foundation and Ministry of Education in Singapore.
We acknowledge useful discussions with M. Paw{\l}owski. T. P. acknowledges discussions with J. Kofler and {\v C}. Brukner, D. K. would like to thank A. Ekert, R. Fazio, V. Scarani and A. Winter for stimulating discussions.

\appendix

\section{Multipartite Scenario}

We generalize the method to the scenario where the system of $N$ qubits is partitioned into $k$ regions, namely $A, B \dots K$ such that there are $N_{k}$ particles in each region. Assume for the moment that all $N_{k}$ are equal to some $n$. Evidently $N = n \times k$. The case where the number of particles in each region is different will be dealt with later. 

Once again, we consider the situation where the local magnetization is measured in each region. The question is then: Does a state $\rho$ of the system exist such that the correlations between the local magnetizations are non-classical? 

We now proceed in a manner analogous to the bipartite scenario. The correlations between local magnetizations read
\begin{eqnarray}
&&\left \langle \mathcal{M}_{\vec n_1} \otimes \dots \otimes \mathcal{M}_{\vec n_k} \right\rangle = 
n^{k} \Tr \Big( (\vec{n}_{1} \cdot \vec{\sigma} \otimes \dots \otimes \vec{n}_{k} \cdot \vec{\sigma}) \rho_{eff}^{R_{1}\dots R_{k}}\Big),
\end{eqnarray}
where the effective state is now a state between $k$ qubits
$$\rho_{eff}^{AB\dots K} = \frac{1}{n^{k}} \sum_{l_{1} \in A} \dots \sum_{l_{k} \in K} \rho_{l_{1} \dots l_{k}}$$
constructed from the $k$-qubit reduced density matrices, $\rho_{l_{1} \dots l_{k}}$, between qubits taken one from each region.
The existence of a LHV model for $k$-qubit correlation measurements in this effective state then implies its existence for the whole quantum state $\rho$.

We use here the results from Ref.\ \cite{zukowski}, in which it was shown that
a set of $2^k$ correlation functions obtained on $k$-qubit state by measuring one of two local observables
admits LHV model if
\begin{equation}
\sum_{i_{1} \dots i_{k} = \{x,y\}} T_{i_1 \dots i_k}^{2} \leq 1,
\end{equation}
where $T_{i_1 \dots i_k} = \Tr(\sigma_{i_1} \otimes \dots \otimes \sigma_{i_k} \rho_{eff}^{AB\dots K} )$
is the correlation function for the orthogonal local directions $\vec x$ and $\vec y$ defined as sum and difference of local measurement settings.
In our case, these correlation functions read
\begin{eqnarray}
&&T_{i_1 \dots i_k} = \frac{1}{n^{k}} \sum_{l_{1} \in R_{1}} \dots \sum_{l_{k} \in R_{k}} T_{i_1 \dots i_k}^{l_{1} \dots l_{k}},
\end{eqnarray}
where $T_{i_1 \dots i_k}^{l_{1} \dots l_{k}}$ gives the correlations between a set of $k$ particles labeled by $l_1 \dots l_k$.
Inserting this expression into the LHV criterion yields
\begin{eqnarray}
&&\sum_{i_{1} \dots i_{k} = 1}^{2} T_{i_1 \dots i_k}^{2} =
\frac{1}{n^{2k}} \sum_{i_{1} \dots i_{k} = 1}^{2} \left(\sum_{l_{1} \dots l_{k}} \sum_{l'_{1} \dots l'_{k}} T_{i_1 \dots i_k}^{l_{1} \dots l_{k}}T_{i_1 \dots i_k}^{l'_{1} \dots l'_{k}}\right).
\label{lhv3}
\end{eqnarray}
We show under which conditions this expression is less than $1$ in order to satisfy the LHV criterion. 

This will be accomplished by showing that the expression above can be written as the sum scalar products between any two of $n^{k}$ vectors each of which has length at most one. Note that the sums over $i_{1} \dots i_{k}$ and the sums over $l_{1} \dots l_{k}$ and $l'_{1} \dots l'_{k}$ result in a total of $2^{k} \times n^{2k}$ terms in the above expression. Hence, each vector that we construct must have a minimum of $2^{k}$ components so that the final expression has magnitude less than $1$. 

From Ref.\ \cite{mono3}, we know that if the components of each vector are averages of mutually anti-commuting observables, the length of the vector is bounded by 1. The task then is to find $n^{k}$ groups of $2^{k}$ correlation functions $T_{i_1 \dots i_k}^{l_{1} \dots l_{k}}$ such that the corresponding observables $\sigma_{i_1} \otimes \dots \otimes \sigma_{i_k}$ mutually anti-commute. This is a generalization of what was done in the bipartite scenario.

We now present a simple algorithm to accomplish this task.
We will first construct one vector of $2^{k}$ components and build the other $n^{k} - 1$ vectors by applying certain modifications to it. 

For simplicity, we shall first construct the vector as a set of $2^{k}$ mutually anti-commuting observables and then replace the observables by the corresponding correlation functions. Note that each component of the vector is a tensor product of $k$ single qubit observables of the type $\sigma_{i_j}$ acting on one of the qubits $l_{j}$ on each region. For each qubit $l_{j}$ in region $J$, these can take only two values, namely $\sigma_1^{l_{j}}$ and $\sigma_2^{l_{j}}$. These two observables clearly anti-commute for given $l_{j}$. Since we need $2^{k}$ mutually anti-commuting observables, a simple and direct solution is to construct a binary tree algorithm which would require $N_{k} = 2^{k-1}$ qubits in a region.

\begin{figure}[!t]
\hspace{-1.5cm}
\vspace{-4cm}
\begin{center}
\includegraphics[width=14cm]{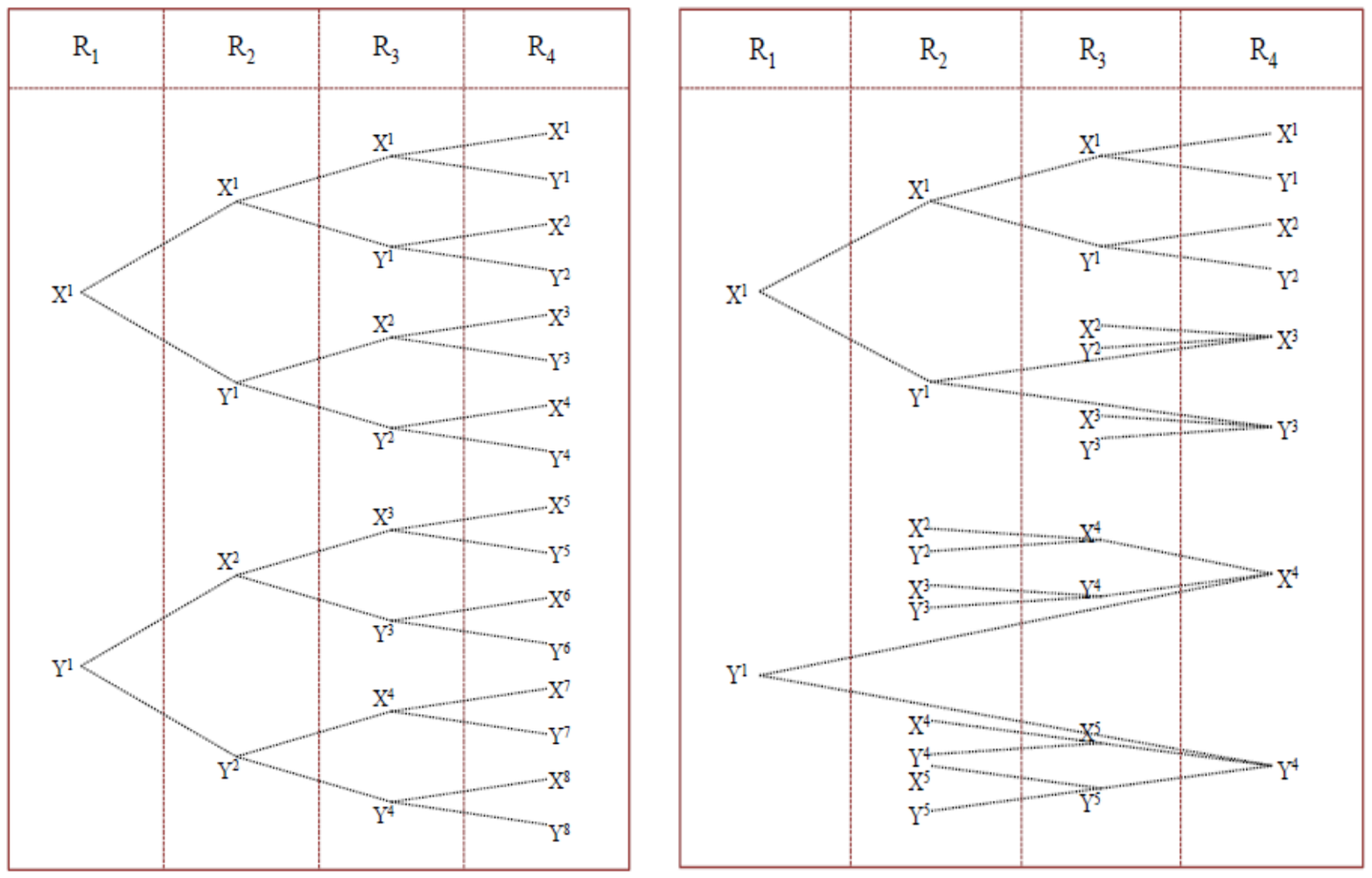}
\end{center}
\vspace{-8.5cm}
\caption{\textbf{Left}: Binary tree construction of $16$ mutually anti-commuting operators for $k = 4$ regions. For convenience $\sigma_{1}$ ($\sigma_{2})$ has been labeled as $X$ ($Y$). Each branch of the tree represents one operator sequence. For instance the top-most branch represents $\sigma_{1}^{1} \otimes \sigma_{1}^{1} \otimes \sigma_{1}^{1} \otimes \sigma_{1}^{1}$. The number of particles in region $J$ is then $n = 2^{j-1}$.
\textbf{Right}: A factor $k$ improvement on $n$ can be obtained by folding the tree at particular operator sequences as explained in the text.}
\label{fig:tree1}
\vspace{-0.3cm}
\end{figure}

Let us first briefly discuss this approach before proceeding to look for improvements. We first list all $2^{k}$ strings $\sigma_1^{l_{1}} \otimes \dots \otimes \sigma_1^{l_{k}}$ to $\sigma_2^{l_{1}} \otimes \dots \otimes \sigma_2^{l_{k}}$. Each of the $n^{k}$ vectors should contain all these strings. The difference between the vectors will lie only in the values of $l_{1} \dots l_{k}$. The simplest approach is to let each $l_{j}$ assume values from 1 to $2^{j-1}$.This can be represented by the tree diagram as shown in Fig.[\ref{fig:tree1}] for $k = 4$. 

The remaining vectors are constructed by simple modifications to the original vector. Two operations are performed:
1. change of $l_{j}$ to $l_{j+m}$ where addition is modulo $m$; and
2. change of $i_{j}$ to $i_{j} + 1$ where addition is modulo 2.

It is straightforward to show that these two operations applied to all operator sequences in a vector preserve the anti-commutation of operators. Moreover, all the $n^{k}$ vectors can be obtained from one vector by applying these two operations.   

Hence, a possible grouping of terms is achieved which ensures that the LHV criterion is satisfied.
The pitfall is that the algorithm is inefficient and needs one of the regions, namely the last one, to have $n = 2^{k-1}$ qubits. Since we assume that all regions contain roughly equal number of qubits, we have $N = 2^{k-1} \times k$. 

A factor $k$ improvement can be obtained by modifying the binary algorithm as we show below. First let us define a function $g(n)$ as the smallest power of 2 that is greater than or equal to $n$. We then define $m$ to be $g(\frac{2^{k-1}}{k-1})$. We then carry out the binary tree algorithm for the first  m operators with the leaf at the $k^{th}$ region. In the second step, we shift the leaf of the tree one region to the left and construct the next $m$ operators again by the binary tree method. Then in the third step, we shift the leaf one region to the left and construct $2m$ operators. In general, in the $j^{th}$ step, we shift the leaf one region to the left and construct $2^{j-2} \times m$ operators by the binary tree method. We carry out this algorithm until $2^{k}$ operators are constructed at which point the binary string is exhausted and no more mutually anti-commuting operators exist.  The algorithm thus describes a binary tree that curls back and equitably distributes the $2^{k}$ operators among the k regions giving at most $n = \sum_{l = 1}^{k} g(\frac{2^{l-1}}{k-1})$ particles per grid. An illustration of this construction for $k = 4$ is given in Fig.\ \ref{fig:tree1}. As before, other vectors are obtained from the first one constructed by applying the two operations previously described. 

A more careful reconstruction of the binary tree is possible to give $N_{k} \geq \left\lceil (\frac{2^{k-2}}{k-1})\right\rceil$. 
This method therefore, assures us that given a sample with $N$ qubits, a division into $k \leq \log_{2}{(N)}$ regions leads to a LHV model for magnetization measurements in two-setting Bell inequalities and more partitions are needed in order to violate such inequalities. It is worth noting that the binary tree method may not be optimal in constructing sets of mutually anti-commuting operators and in actual fact, the dependence of $N$ on $k$ may be polynomial rather than exponential. One further point to be noted is that when the number of qubits in each region is different, $n = \left\lceil (\frac{2^{k-2}}{k-1})\right\rceil$ represents the minimum number of qubits in any region that ensures the LHV model. It can be shown that the two operations described yield all vectors of mutually anti-commuting operators in this scenario as well.

\end{document}